\documentclass[aps,pra,reprint,superscriptaddress,nobibnotes]{revtex4-1}

\usepackage{graphicx}
\usepackage{bm}
\usepackage{bbold}
\usepackage{amssymb}
\usepackage{amsmath}

\usepackage{color}

\renewcommand{\vec}[1]{{\mathbf #1}}

\begin{document}

\title{Intensity of waves inside a strongly disordered medium}

\author{S.E. Skipetrov}
\email[]{Sergey.Skipetrov@lpmmc.cnrs.fr}
\affiliation{Univ. Grenoble Alpes, CNRS, LPMMC, 38000 Grenoble, France}

\author{I.M. Sokolov}
\email[]{ims@is12093.spb.edu}
\affiliation{Department of Theoretical Physics, Peter the Great St. Petersburg Polytechnic University, 195251 St. Petersburg, Russia}

\date{\today}

\begin{abstract}
Anderson localization does not lead to an exponential decay of intensity of an incident wave with the depth inside a strongly disordered three-dimensional medium. Instead, the average intensity is roughly constant in the first half of a disordered slab, sharply drops in a narrow region in the middle of the sample, and then remains low in the second half of the sample. A universal, scale-free spatial distribution of average intensity is found at mobility edges where the intensity exhibits strong sample-to-sample fluctuations. Our numerical simulations allow us to discriminate between two competing local diffusion theories of Anderson localization and to pinpoint a deficiency of the self-consistent theory.
\end{abstract}

\maketitle

Studies of wave propagation in disordered media mainly focus on the scattering problem in which one is interested in determining a relation between incident and scattered waves outside the disordered sample and often even in the far field of it \cite{sheng06,akkermans07}. Transmission and reflection coefficients of disordered media have been extensively studied in this context, including their statistics and correlations \cite{akkermans07}. Scattered waves outside the medium are not only easier to measure, they are also relevant for understanding practically important quantities, such as the electrical conductance of metals \cite{dugdale95} or the whiteness of paints \cite{palmer89}, as well as for developing applications for complex material \cite{scheffold03} or biological tissue \cite{durduran10} sensing, imaging through opaque, turbid media \cite{katz14}, or cryptography \cite{goorden14}. In contrast, the spatial distribution of wave intensity \textit{inside} a disordered medium has attracted much less attention even though it is important for such prospective applications of disordered materials as light harvesting in solar cells \cite{vynck12}, random lasing \cite{wiersma08}, optical frequency conversion \cite{fischer06} or photoacoustic tomography \cite{wang12}. For three-dimensional (3D) media we know that the average intensity exhibits diffusive behavior for weak disorder and hence, in the absence of absorption, decays linearly with the depth inside a disordered layer (slab) illuminated by a plane wave \cite{sheng06,akkermans07}. However, nothing is known at the moment about the way in which this linear behavior is modified when the disorder becomes strong enough for reaching a critical point of the Anderson localization transition (a mobility edge) and crossing it to enter the Anderson localization regime \cite{anderson58,lagendijk09}.

The spatial distribution of the average wave intensity $\langle I(\vec{r}) \rangle$ inside a strongly disordered medium of length $L$ illuminated by a monochromatic wave has been studied theoretically for a one-dimensional (1D) medium \cite{gazaryan69,lang73,abram79,mello16} and for a quasi-one dimensional (quasi-1D) waveguide \cite{zhao13,tiggelen17,cheng17}. In both cases, the behavior of $\langle I(\vec{r}) \rangle = \langle I(z) \rangle$ differs from a simple exponential decay with the distance $z$ from the sample boundary. This suggests that the exponential decay of eigenmodes in space does not directly map to the exponential decay of the average intensity. Instead, $\langle I(z) \rangle$ exhibits a step-like shape, first remaining virtually constant with $z$, then dropping sharply in a narrow region around the middle of the disordered sample $z = L/2$, and finally remaining low for $L/2 < z < L$. A tendency towards such a behavior has been experimentally observed by Yamilov et al. in two-dimensional (2D) quasi-1D waveguides \cite{yamilov14}.

In this Letter we use ab initio numerical simulations of wave scattering in large 3D ensembles of point scatterers and the local diffusion theories of Anderson localization to discover two important results. First, we show that the behavior that was previously found for $\langle I(z) \rangle$ in 1D and quasi-1D samples, generalizes to 3D slabs, provided that the disorder is strong enough for reaching Anderson localization. Two competing local diffusion theories---the self-consistent (SC) theory of Anderson localization and the supersymmetric (SUSY) field theory---yield analytic expressions for $\langle I(z) \rangle$ as a function of $z/L$ that are parameterized by a single parameter $L/\xi$, where $L$ is the slab thickness and $\xi$ is the localization length. Second, we compute $\langle I(z) \rangle$ at a mobility edge, i.e. in the critical regime that does not exist in low-dimensional systems. Analytic expressions for $\langle I(z) \rangle$ following from SC and SUSY theories become scale-independent for $L$ much exceeding the mean free path $\ell$. By repeating calculations for light scattering by atoms in a strong magnetic field we demonstrate that our results are universal and hold beyond the scalar wave model. This completes the palette of behaviors expected for $\langle I(z) \rangle$ for any disorder strength, any dimensionality of space, and for both scalar and vector waves. Comparison of SC theory with numerical simulations and SUSY theory confirms its validity at the mobility edge but reveals its deficiency in the Anderson localization regime. Understanding limitations of SC theory is important in view of its applications for interpretation of 3D acoustic \cite{cobus16,cobus18} and cold-atom \cite{jendr12} experiments as well as of large-scale numerical simulations of light localization \cite{haberko18}.

We consider a monochromatic plane wave $\psi_0(\vec{r}) = \exp(i k z)$ incident at $z = 0$ on a disordered sample (slab) confined between the planes $z = 0$ and $z = L$ and having a shape of a cylinder of length (thickness) $L$, radius $R \gg L$ and volume $V = \pi R^2 L$. We denote the frequency of the wave by $\omega$ and its wave number by $k = \omega/c$, where $c$ is the speed of the wave in the homogeneous medium by which the sample is surrounded. Our point-scatterer model assumes that the sample is simply an ensemble of $N \gg 1$ identical resonant point scatterers with a polarizability $\alpha(\omega) = -(\Gamma_0/2)/(\omega - \omega_0 + i \Gamma_0/2)$ located at random positions $\{ \vec{r}_m \}$, $m = 1, \ldots, N$, inside the slab. The resonance width $\Gamma_0$ is assumed to be much smaller than the resonance frequency $\omega_0$ of an individual scatterer. A vector $\bm{\psi} = [\psi(\vec{r}_1), \ldots, \psi(\vec{r}_N)]^{T}$ of wave amplitudes at scatterer positions obeys \cite{foldy45,lax51}
\begin{eqnarray}
\bm{\psi} = \bm{\psi}_0 + \alpha(\omega) \left[ {\hat G}(\omega) - i \mathbb{1} \right] \bm{\psi},
\label{foldylax}
\end{eqnarray}
where $\bm{\psi}_0 = [\psi_0(\vec{r}_1), \ldots, \psi_0(\vec{r}_N)]^{T}$ and
\begin{eqnarray}
G_{mn}(\omega) = i \delta_{mn} + \left( 1-\delta_{mn} \right) \frac{\exp(i k |\vec{r}_m - \vec{r}_n|)}{k |\vec{r}_m - \vec{r}_n|}.
\label{green}
\end{eqnarray}
The solution of Eq.\ (\ref{foldylax}) reads
\begin{eqnarray}
\bm{\psi} = \left( \mathbb{1} - \alpha(\omega) \left[ {\hat G}(\omega) - i \mathbb{1} \right] \right)^{-1} \bm{\psi}_0.
\label{sol}
\end{eqnarray}
We compute the average intensity $\langle I(\vec{r}) \rangle$ inside the sample by averaging $| \psi(\vec{r}_m) |^2$ over all $\vec{r}_m$ inside a small volume around $\vec{r}$ and over many (up to $5 \times 10^5$) random and statistically independent scatterer configurations $\{ \vec{r}_m \}$. In addition, $\langle I(\vec{r}) \rangle$ is averaged over a sufficiently large circular area of radius $R_1$ around the sample axis ($1/k \ll R_1 < R$) in order to obtain $\langle I(z) \rangle$ which is independent of $\vec{r}_{\perp} = \{x, y \}$ and mimics the average intensity in a disordered slab of infinite transverse extent $R \to \infty$.

We have extensively studied Anderson localization in the model defined by Eqs.\ (\ref{foldylax}--\ref{sol}) in our previous works \cite{skip16prb,skip18ir}. In particular, we have found that spatially localized modes appear in a narrow frequency band between two density-dependent mobility edges $\omega_c^{\mathrm{I}} = \omega_c^{\mathrm{I}}(\rho/k_0^3)$ and $\omega_c^{\mathrm{II}} = \omega_c^{\mathrm{II}}(\rho/k_0^3)$ for scatterer number densities $\rho = N/V$ exceeding a critical value $\rho_c \simeq k_0^3/4\pi$,
where $k_0 = \omega_0/c$. We will use these previous results to study $\langle I(z) \rangle$ in the localized regime by choosing the frequency $\omega \in (\omega_c^{\mathrm{I}}, \omega_c^{\mathrm{II}})$ and in the critical regime for $\omega = \omega_c^{\mathrm{I}}$ or $\omega = \omega_c^{\mathrm{II}}$.

\begin{figure}[t]
\hspace*{-0.55cm}
\includegraphics[width=1\columnwidth]{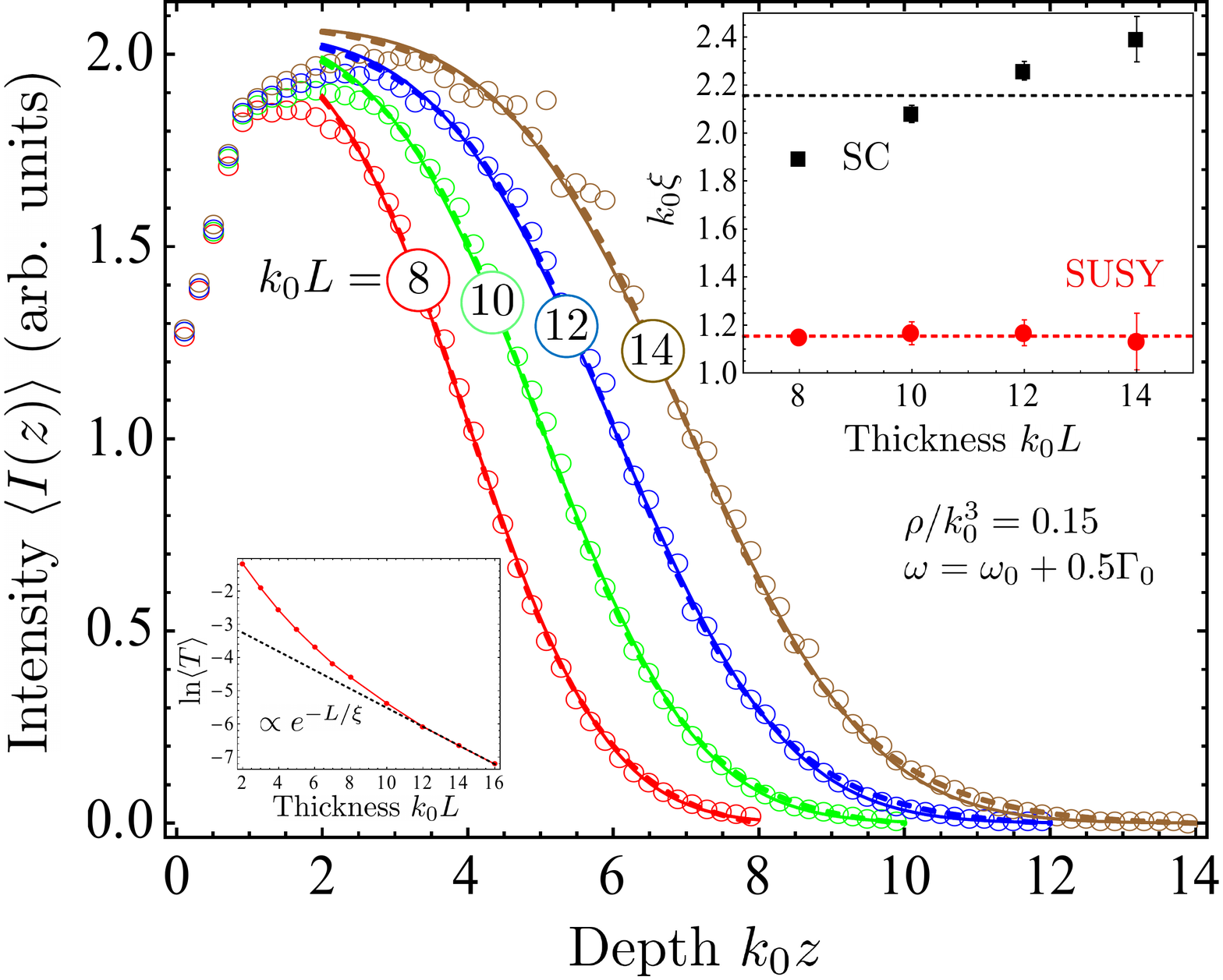}
\caption{\label{fig_loc}
Spatial distributions of the average wave intensity inside slabs of disordered medium of different thicknesses $k_0 L = 8$--14. Symbols correspond to the point-scatterer model (\ref{foldylax}) with $\omega$ in between the two mobility edges $\omega_c^{\mathrm{I}} = \omega_0 + 0.256 \Gamma_0$ and $\omega_c^{\mathrm{II}} = \omega_0 + 0.935 \Gamma_0$ for $\rho/k_0^3 = 0.15$ \cite{skip18ir}, $k_0 R = 20$ and $k_0 R_1 = 10$. Almost coinciding dashed and solid lines show fits of SC (\ref{ilocsc}) and SUSY (\ref{ilocsusy}) theories to the point-scatterer data for $k_0 z > 2$ with the localization length $\xi$ as a free parameter. The upper inset shows the best-fit values of $\xi$ following from SC (black squares) and SUSY (red circles) theories. Dashed lines show average values of $\xi$. The lower inset shows the average transmission through the slab as a function of slab thickness.}
\end{figure}

\begin{figure*}[t]
\includegraphics[width=1\columnwidth]{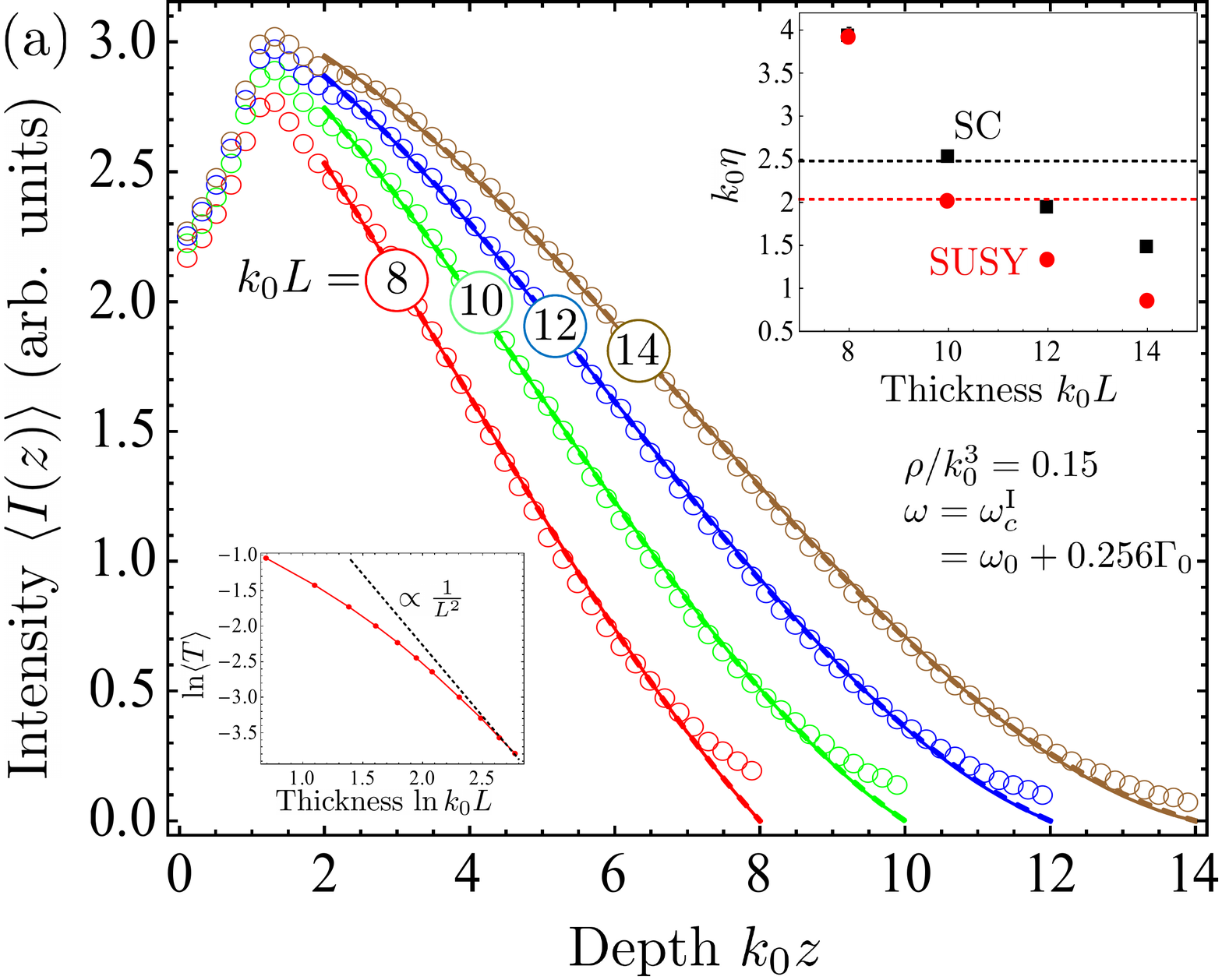}
\includegraphics[width=1\columnwidth]{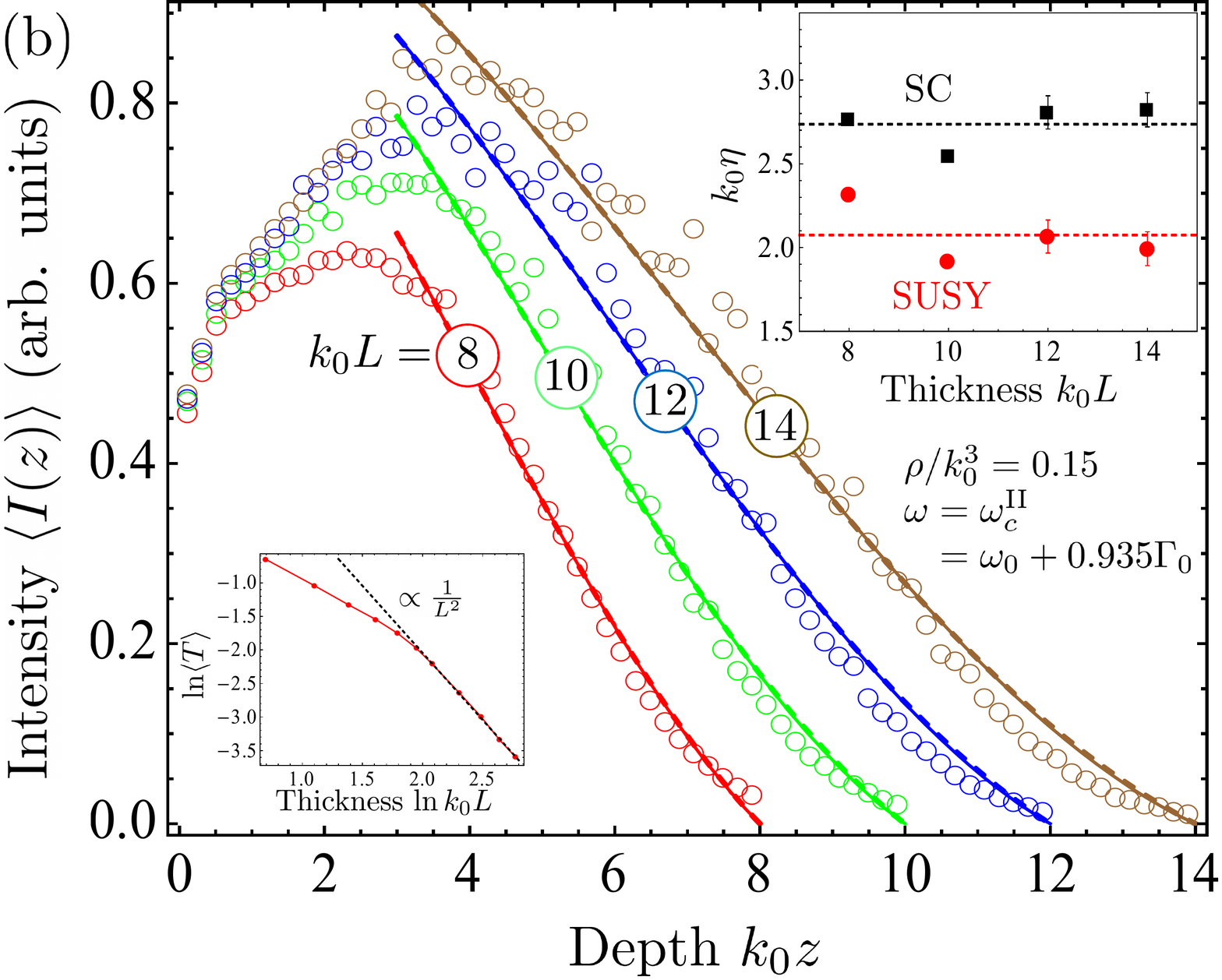}
\caption{\label{fig_me}
Same as Fig.\ \ref{fig_loc} but at the low- (a) and high-frequency (b) mobility edges $\omega = \omega_c^{\mathrm{I}}$ and $\omega = \omega_c^{\mathrm{II}}$, respectively, and for $k_0 R = 25$. Dashed and solid lines show, respectively, SC and SUSY theory fits to numerical results with the decay length $\eta$ as a free fit parameter. The fits were performed for $k_0 z > 2$ (a) or $k_0 z > 3$ (b). The upper insets show the best-fit values of $\eta$ for SC (black squares) and SUSY (red circles) models, with average values of $\eta$ represented by dashed lines. The lower insets show the average transmission through the slab as a function of slab thickness.}
\end{figure*}

The results of the point-scatterer model (\ref{foldylax}--\ref{sol}) will be compared to two competing local diffusion theories of Anderson localization \cite{tiggelen00,cherroret08,tian08,tian10,tian13}. In these theories, the average intensity of a wave $\langle I(\vec{r}) \rangle$ obeys a diffusion equation with a position-dependent diffusivity $D(\vec{r})$:
\begin{eqnarray}
&& -\boldsymbol{\nabla} \cdot D(\vec{r}) \boldsymbol{\nabla}
\langle I(\vec{r}) \rangle = S(\vec{r}),
\label{dif}
\end{eqnarray}
where $S(\vec{r})$ describes the distribution of wave sources in the medium.
In 3D, the position dependence of $D(\vec{r})$ in Eq.\ (\ref{dif}) arises only for strong disorder and can be found in two different ways. First, SC theory of localization \cite{vollhardt80,vollhardt92,tiggelen00} yields $D(\vec{r})$ determined self-consistently via the return probability $P(\vec{r}, \vec{r}'= \vec{r})$ found as a solution of Eq.\ (\ref{dif}) with $S(\vec{r}) = \delta(\vec{r}-\vec{r}')$ and an appropriate cut-off procedure to regularize the unphysical divergence of the solution for $\vec{r}'= \vec{r}$ \cite{tiggelen00,cherroret08}:
\begin{eqnarray}
\frac{1}{D(\vec{r})} =  \dfrac{1}{D_B}+\dfrac{12\pi}{K^2\ell}P(\vec{r},\vec{r}),
\label{sc2}
\end{eqnarray}
where $D_B$ is the bare value of $D$ in the absence of localization effects and  $K$ is the effective wave number in the disordered medium.
In a slab, Eqs.\ (\ref{dif}) and (\ref{sc2}) should be solved with appropriate boundary conditions for $P(\vec{r},\vec{r}')$ \cite{sm}.
A second approach is based on field-theoretic, SUSY methods and has been mainly developed for 1D and quasi-1D media \cite{tian08,tian10,tian13}. It does not provide a simple microscopic expression or an equation for $D(\vec{r})$ that would hold for any sample geometry, but it yields a scaling relation between $D_{\infty}(z)$ in the semi-infinite medium and $D(z)$ in a slab of finite thickness $L$ \cite{tian10}.

Interestingly enough, both SC and SUSY theories yield $D_{\infty}(z) = D(0) \exp(-2 z/\xi)$ in the Anderson localization regime ($K\ell < 1$) \cite{note1}, but solutions for the slab geometry differ. SC theory yields a result that for $L \gg \xi$ is well described by an interpolation formula $D_{\mathrm{SC}}(z) = [D_{\infty}(z)^{1/2} + D_{\infty}(L-z)^{1/2}]^2$ \cite{sm}. The SUSY approach yields a different result: $D_{\mathrm{SUSY}}(z) = D_{\infty}[z(L-z)/L]$ \cite{tian10}. In both cases, flux conservation implies that
the diffusive flux $J_{\mathrm{dif}}$ given by the Fick's law
$J_{\mathrm{dif}} = -D(z) \partial \langle I(z) \rangle/\partial z$ is independent of $z$ for $z \gg \ell$.
Integrating the Fick's law yields
\begin{eqnarray}
\langle I(z) \rangle = I_0 - J_{\mathrm{dif}} \int_0^z \frac{dz'}{D(z')}
\;\; \text{for~} z \gg \ell,
\label{igeneral}
\end{eqnarray}
where the precise value of $I_0$ depends on the details of conversion of the incident plane wave into diffuse radiation near the sample surface $z = 0$. Supplemented with a boundary condition $\langle I(L) \rangle = 0$ \cite{akkermans07}, Eq.\ (\ref{igeneral}) yields
\begin{eqnarray}
\langle I(z) \rangle_{\mathrm{SC}} &=& I_0
\frac{\sinh[(L-z)/\xi]}{\sinh(z/\xi) + \sinh[(L-z)/\xi]},
\label{ilocsc}
\\
\langle I(z) \rangle_{\mathrm{SUSY}} &=& \frac{I_0}{2}
\left\{ 1 +
\frac{\mathrm{erf} \left[ (L - 2 z)/\sqrt{2 L \xi} \right]}{\mathrm{erf}( \sqrt{L/2 \xi} )}
\right\},\;\;\;\;\;
\label{ilocsusy}
\end{eqnarray}
where $\mathrm{erf}(x)$ is the error function.

The point-scatterer model and SC and SUSY theories yield consistent results for the distribution of the average intensity $\langle I(z) \rangle$ inside the disordered slab in the localized regime. As we see from Fig.\ \ref{fig_loc}, $\langle I(z) \rangle$ does not decay exponentially with $z$ as one could expect from naive considerations, but instead exhibits a rapid drop near the middle of the slab, while varying much slower near its boundaries. Such a behavior is similar to that found previously in 1D \cite{gazaryan69,lang73,abram79,mello16,zhao13} and quasi-1D \cite{tiggelen17,cheng17} media.

Even though both SC and SUSY theories provide good and, in fact, hardly distinguishable fits to the numerical data, only SUSY model consistently yields the same (within error bars) best-fit values of $\xi$ for different $L$ as we show in the inset of Fig.\ \ref{fig_loc}.
The underlying problem of SC model is best demonstrated by computing the width $\delta$ of the spatial region in which the average intensity changes rapidly near the middle of the sample:
\begin{eqnarray}
\delta = \left. \left[ -\frac{1}{\langle I(z) \rangle} \frac{\partial}{\partial z} \langle I(z) \rangle
\right]^{-1}\right|_{z=L/2}.
\label{eta}
\end{eqnarray}
We find $\delta_{\mathrm{SC}} = \xi$ and $\delta_{\mathrm{SUSY}} = \sqrt{(\pi/8) L \xi}$ for $L \gg \xi$, which predict different scalings of $\delta$ with $L$. The need for different values of $\xi$ to fit the numerical data corresponding to different $L$ with SC theory signals that the scaling that it predicts for $\delta$ is wrong. In contrast, SUSY theory yields the correct scaling for $\delta$ and describes the data in Fig.\ \ref{fig_loc} with a single value of $\xi$ for all $L$.


An interesting regime that is not accessible in low-dimensional systems is the critical one. In order to study it in the framework of the point-scatterer model (\ref{foldylax}), we choose the frequency of the wave $\omega$ exactly at one of the the mobility edges $\omega_c^{\mathrm{I, II}}$ determined in Ref.\ \cite{skip18ir}. The resulting spatial distributions of $\langle I(z) \rangle$ are shown in Fig.\ \ref{fig_me} by symbols. To study the critical regime using the local diffusion theories, we note that for a semi-infinite medium ($L \to \infty$) one finds $D(z) = D_{\infty}(z) \simeq D(0)/(1 + z/\eta)$ \cite{tiggelen00} with a decay length $\eta \sim \ell$. For a slab of finite thickness $L$, the results of SC theory may be nicely interpolated by $D(z) = [D_{\infty}(z)^2 + D_{\infty}(L-z)^2]^{1/2}$ \cite{sm}, whereas
another option is to extrapolate the relation $D_{\mathrm{SUSY}}(z) = D_{\infty}[z(L-z)/L]$ \cite{tian10,tian13} to the mobility edge. Proceeding in the same way as for deriving Eqs.\ (\ref{ilocsc}) and (\ref{ilocsusy}), we obtain expressions for $\langle I(z) \rangle$ that depend on $z/L$ and $L/\eta$. The full expression following from SC theory is quite cumbersome and we reproduce it elsewhere \cite{sm} whereas the SUSY result is simpler:
\begin{eqnarray}
\langle I(z) \rangle_{\mathrm{SUSY}} &=& I_0
\left( 1 - \frac{z}{L} \right)
\left[ 1 + \frac{\frac{z}{L} (1 - 2 \frac{z}{L})}{1 + 6 \frac{\eta}{L}} \right].\;\;\;\;
\label{imesusy}
\end{eqnarray}
Comparison of these results with numerical simulations of the model (\ref{foldylax}) is shown in Fig.\ \ref{fig_me}. The agreement is less striking than in the localized regime but it improves when $L$ increases. In addition, variations of the best-fit $\eta$ with $L$ are similar for SC and SUSY theories but differ from SC theory expectations \cite{sm}. Universal, parameter-free intensity distributions follow in the limit of $L \gg \eta$:
\begin{eqnarray}
&&\langle I(z) \rangle_{\mathrm{SC}} = \frac{I_0}{2}
\left\{
\vphantom{\frac{1}{2}}
1
\right.
\label{imesc0}
\\
&& + \left. \frac{3 \sqrt{2} \mathrm{arcsinh}(1 - 2\frac{z}{L})- 2 (1 - 2\frac{z}{L}) \sqrt{1 - 2\frac{z}{L}(1-\frac{z}{L})}}{
3\sqrt{2}\, \mathrm{arcsinh}(1) - 2}
\right\},
\nonumber
\\
&&\langle I(z) \rangle_{\mathrm{SUSY}} = I_0
\left( 1 - \frac{z}{L} \right)^2
\left( 1 + 2 \frac{z}{L} \right).
\label{imesusy0}
\end{eqnarray}
These two expressions are very close when plotted as functions of $z/L$.
Their lack of any characteristic length scale can be seen as a consequence of the fractal character of critical eigenmodes \cite{evers08}. By analogy with other models of Anderson localization \cite{evers08,rodriguez09}, we expect the critical eigenmodes of our model defined by Eqs.\ (\ref{foldylax}--\ref{sol}) to be \textit{multifractal}, but evidencing this would require analysis of higher-order statistical moments and spatial correlations of intensity in addition to the study of its average value.


\begin{figure}[t]
\hspace*{-8.5mm}
\includegraphics[width=1.1\columnwidth]{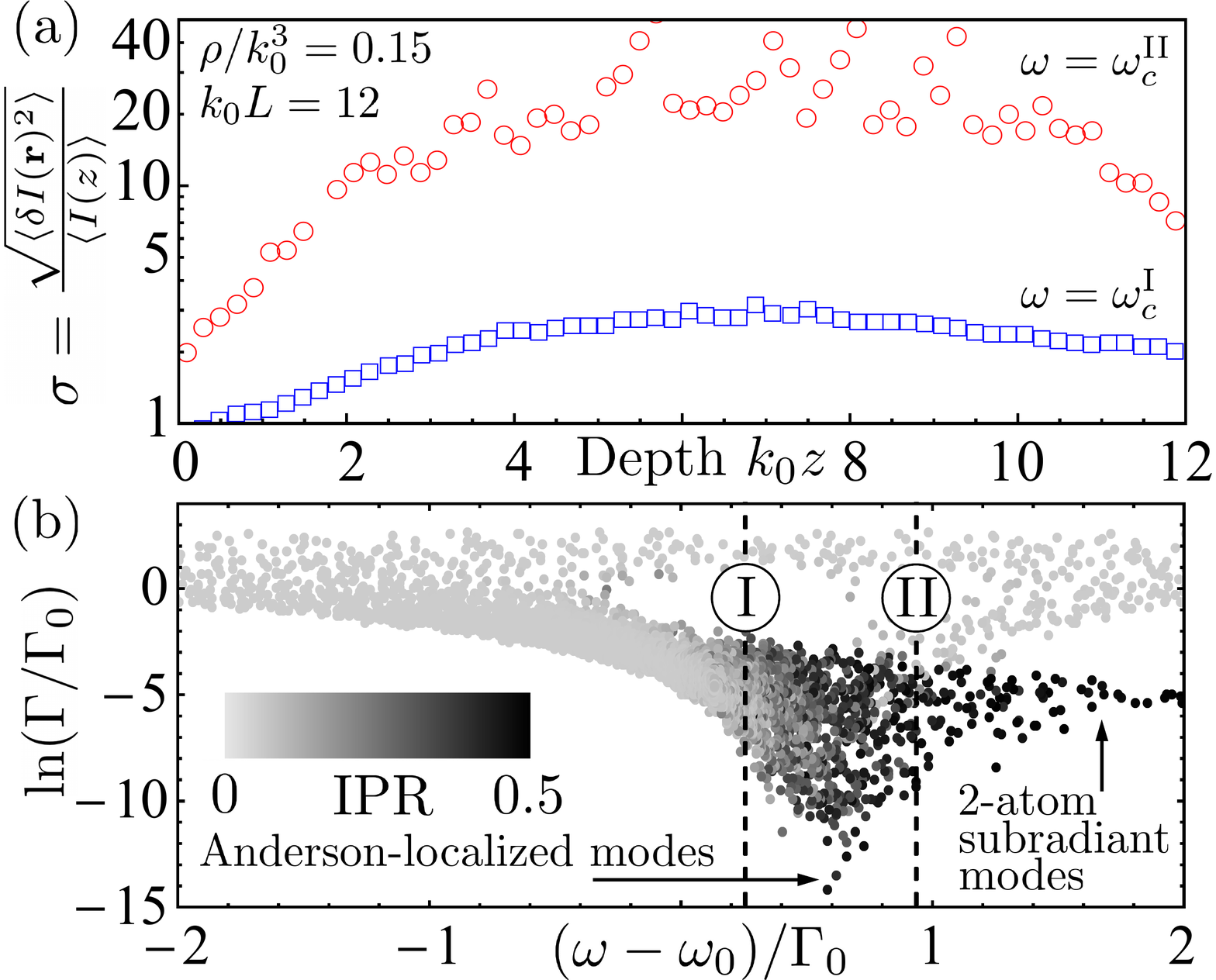}
\caption{\label{fig_fluct}
(a) Relative fluctuations of intensity at the two mobility edges for a slab of thickness $k_0 L = 12$ and all other parameters as in Fig.\ \ref{fig_me}.
(b) Grayscale plot of IPR of quasimodes for a representative random configuration of scatterers. Each point corresponds to a quasimode of frequency $\omega = \omega_0 - (\Gamma_0/2) \mathrm{Re}\Lambda$ and decay rate $\Gamma = \Gamma_0 \mathrm{Im}\Lambda$, where $\Lambda$ is an eigenvalue of ${\hat G(\omega_0)}$.  The gray scale of the point reflects the IPR of the corresponding eigenvector. Dashed lines show the two mobility edges.
}
\end{figure}

Comparison of results corresponding to the two mobility edges $\omega = \omega_c^{\mathrm{I}}$ and $\omega = \omega_c^{\mathrm{II}}$ suggests that the behaviors of our point-scatterer model at these frequencies are quite different. First, the mean free path $\ell$ can be estimated as a position of the maximum of $\langle I(z) \rangle$ in Figs.\ \ref{fig_me}(a) or (b) and turns out to be considerably larger at the second, high-frequency mobility edge. As a consequence, the results presented in Fig.\ \ref{fig_me}(b) correspond to shorter optical thicknesses $L/\ell$ than the data in Fig.\ \ref{fig_me}(a). Second, the sample-to-sample fluctuations of intensity at the second mobility edge are much stronger than at the first one. This is illustrated in Fig.\ \ref{fig_fluct}(a) where we show the relative intensity fluctuation $\sigma = \sqrt{\langle \delta I(\vec{r})^2 \rangle}/\langle I(z) \rangle$, where $\delta I(\vec{r}) = I(\vec{r}) - \langle I(z) \rangle$, as a function of $z$ for $k_0 L = 12$.
We attribute this difference to subradiant states localized on pairs of closely located scatterers and surviving multiple scattering only for small interatomic distances and hence large frequency shifts $(\omega-\omega_0)/\Gamma_0$ \cite{skip16prb}. Figure\ \ref{fig_fluct}(b) shows the inverse participation ratio $\mathrm{IPR}_n = \sum_m |\psi_n(\vec{r}_m)|^4$ of eigenvectors $\bm{\psi}_n$ of the matrix ${\hat G}(\omega_0)$ (quasimodes) as a function of their frequencies $\omega$ and decay rates $\Gamma$. Note that whereas the low-frequency mobility edge I defines a sharp transition between extended states for $\omega < \omega_c^{\mathrm{I}}$ (low IPR, light grey points) and localized states for $\omega_c^{\mathrm{I}} < \omega < \omega_c^{\mathrm{II}}$ (high IPR, dark grey and black points), there are localized states on both sides from the mobility edge II. However, the physical origin of quasimode localization is different for $\omega_c^{\mathrm{I}} < \omega < \omega_c^{\mathrm{II}}$ (localization due to strong scattering appearing only for $N \gg 1$ and $\rho > \rho_c$) and $\omega > \omega_c^{\mathrm{II}}$ (localization that exist for any $N \geq 2$ and any $\rho$). This difference is manifest in the scaling properties of quasimode properties with sample size \cite{skip16prb}. Its link with existence of two-atom subradiant states is further confirmed by a more detailed analysis \cite{sm}.


In conclusion, we have found analytic formulas for the spatial distribution of average wave intensity $\langle I(z) \rangle$ inside a thick 3D slab of strongly disordered medium illuminated by a monochromatic plane wave. In the Anderson localization regime, $\langle I(z) \rangle$ exhibits a step-like shape and drops sharply within a region of width $\delta \sim \sqrt{L \xi}$ in the middle of the sample. At a mobility edge, $\langle I(z) \rangle$ takes a universal, parameter-free shape as a function of $z/L$. Comparison of ab initio numerical simulations with local diffusion theories allowed us to reveal a deficiency of SC theory for description of Anderson localization in 3D.
A realistic physical system in which Anderson localization of light can be observed is a large ensemble of cold atoms in a strong magnetic field \cite{skip15,skip18}. Repeating all the calculations presented above for this system yields very similar
results \cite{sm}.
In a cloud of two-level cold atoms, intensity of light is proportional to the population of the excited state and therefore its spatial distribution can be imaged by the so-called diffraction-contrast imaging \cite{turner05} allowing for state-selective imaging of atoms \cite{sheludko08}, by monitoring a slow spontaneous decay of the excited state to a third, auxiliary level, or by probing the excited level by a weak probe beam resonant with a transition to a higher-energy state. In a dielectric disordered system, spatial distribution of optical intensity can be imaged by optoacoustic methods \cite{karabutov99}.

Numerical calculations of the spatial distributions of average intensity and of the transmission coefficients were carried out with the financial support of the Russian Science Foundation (Project No. 17-12-01085). IMS acknowledges the hospitality of the LPMMC where a part of this work has been performed with a financial support of the Centre de Physique Th\'{e}orique de Grenoble-Alpes (CPTGA).


\renewcommand{\vec}[1]{{\mathbf #1}}
\renewcommand{\thefigure}{S\arabic{figure}}
\renewcommand{\theequation}{S\arabic{equation}}

\bibliographystyle{apsrev4-1}
\renewcommand*{\citenumfont}[1]{S#1}
\renewcommand*{\bibnumfmt}[1]{[S#1]}

\setcounter{equation}{0}
\setcounter{figure}{0}


\onecolumngrid

\newpage
\begin{center}
\Large{\bf{Supplemental Material}}
\end{center}

\begin{center}
\parbox{0.75\textwidth}{\small We justify the approximate interpolation formulas for the position-dependent diffusivity following from SC theory of localization and give the full expression for the position dependence of the average intensity in a disordered slab following from SC theory at a mobility edge. We calculate the average intensity of light in a dense ensemble of two-level atoms subjected to a static magnetic field and demonstrate that it can be described by the analytical formulas that we have derived.
Finally, we discuss the relation between subradiant states localized on pairs of closely located atoms and intensity fluctuations at mobility edges.}
\end{center}

\twocolumngrid

\noindent
\textbf{Position-dependent diffusivity from SC theory of localization.}
Self-consistent (SC) theory of localization was formulated by Vollhardt and W\"{o}lfle \cite{vollhardt80sm,vollhardt92sm,wolfle10sm}. Later on, Van Tiggelen \textit{et al.} extended it by allowing for a position dependence of the renormalized diffusivity $D$ \cite{tiggelen00sm}. This position dependence has been rigourously justified \cite{cherroret08sm,tian08sm}. For the stationary case (i.e., no time dependence), SC theory reduces to two self-consistent equations:
\begin{eqnarray}
&& -\boldsymbol{\nabla} \cdot D(\vec{r}) \boldsymbol{\nabla}
P(\vec{r}, \vec{r}') = \delta(\vec{r}-\vec{r}'),
\label{sc1}
\\
&&\frac{1}{D(\vec{r})} =  \dfrac{1}{D_B}+\dfrac{12\pi}{K^2\ell}P(\vec{r},\vec{r}),
\label{sc2}
\end{eqnarray}
where $D_B$ is the diffusivity in the absence of localization effects, $K$ is the effective wave number in the disordered medium, and $\ell$ is the mean free path.

The return probability $P(\vec{r},\vec{r})$ plays the key role in Eqs.\ (\ref{sc1}) and (\ref{sc2}). Because of the divergence $\propto 1/|\vec{r} - \vec{r}'|$ of $P(\vec{r}, \vec{r}')$ following from Eq.\ (\ref{sc1}), a cutoff is needed to regularize $P(\vec{r}, \vec{r})$ entering Eq.\ (\ref{sc2}). We implement the cutoff in the Fourier space and, anticipating the slab geometry to which we will apply Eqs.\ (\ref{sc1}) and (\ref{sc2}) in the following, only for the component $\vec{q}_{\perp}$ of the wavevector $\vec{q} = \{ \vec{q}_{\perp}, q_z \}$:
\begin{eqnarray}
P(\vec{r},\vec{r}) = \frac{1}{2 \pi} \int_0^{q_{\perp}^{\mathrm{max}}}
d q_ {\perp} q_ {\perp} {\hat P}(q_ {\perp}, z = z^{\prime}),
\label{ft2}
\end{eqnarray}
where
\begin{eqnarray}
{\hat P}(q_ {\perp}, z = z^{\prime}) &=& \int d^2 \vec{r}_{\perp}
P(\vec{r} = \{ \vec{r}_{\perp},z \}, \vec{r}^{\prime} = \{ \vec{r}_{\perp}^{\prime},z \})
\nonumber \\
&\times& \exp\left[-i \vec{q}_{\perp} (\vec{r}_{\perp}-\vec{r}_{\perp}^{\prime}) \right]
\label{ft1}
\end{eqnarray}
and $\vec{r}_{\perp} = \{ x, y \}$.

\begin{figure}[t]
\includegraphics[width=\columnwidth]{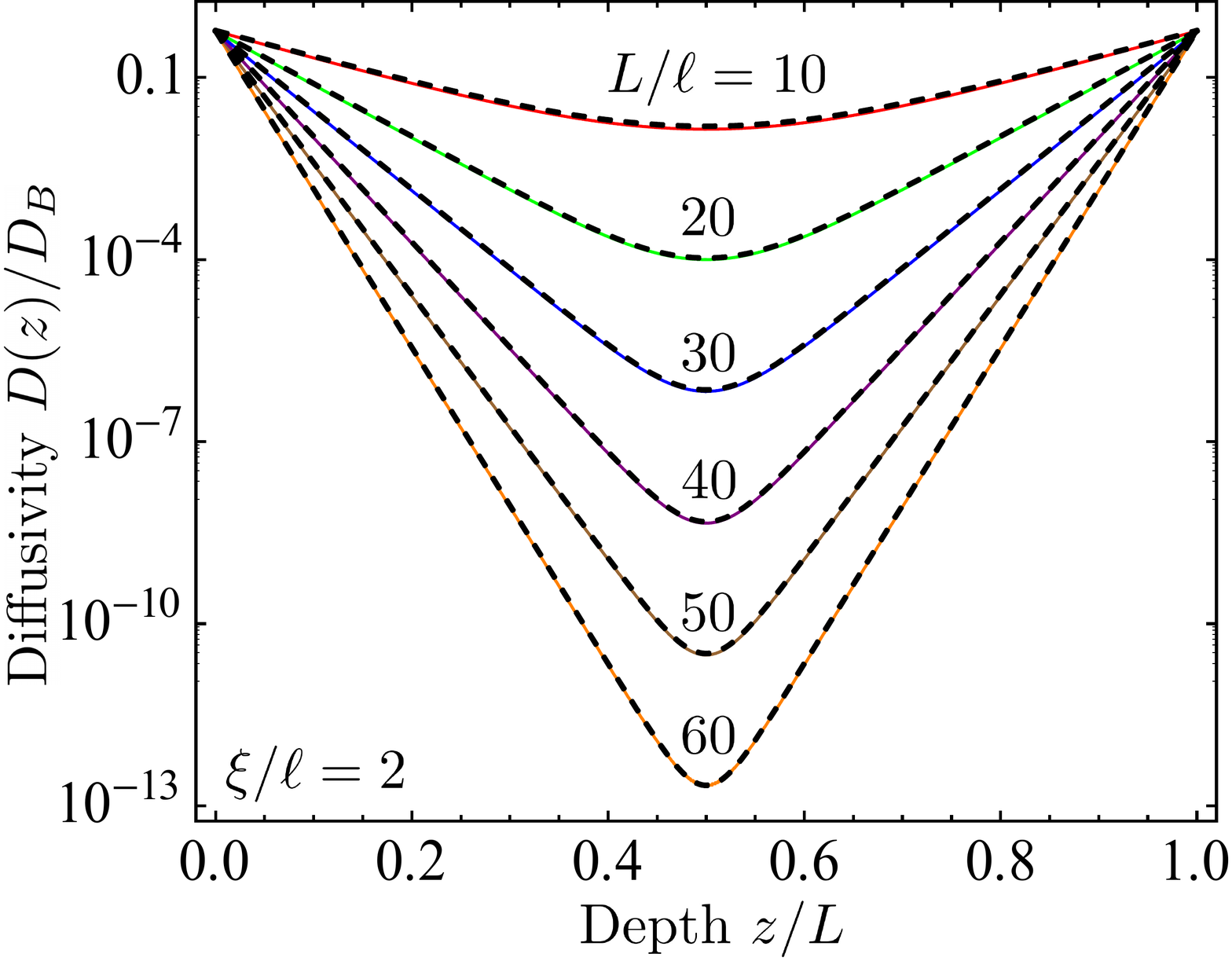}
\caption{\label{fig_dz_loc}
Position dependence of the diffusivity $D(z)$ in disordered slabs of different thicknesses $L$ in the localized regime. Solid lines show the numerical solutions of Eqs.\ (\ref{sc1})--(\ref{bc}). Dashed lines show Eq.\ (\ref{dzloc}). There are no free fit parameters for the dashed lines.}
\end{figure}

The cutoff $q_{\perp}^{\mathrm{max}}$ determines the critical value $(K\ell)_c$ of the Ioffe-Regel parameter $K\ell$ for which $D$ vanishes and the Anderson localization transition takes place in the infinite disordered medium. We set $q_{\perp}^{\mathrm{max}} = \pi/6\ell$ corresponding to $(K\ell)_c = 1$. Then the solution of Eqs.\ (\ref{sc1}) and (\ref{sc2}) decays exponentially in space for $K\ell < 1$: $P(\vec{r}, \vec{r}^{\prime}) = \exp(-|\vec{r}-\vec{r}^{\prime}|/\xi)/(4 \pi \xi^2 |\vec{r}-\vec{r}^{\prime}|)$, with a localization length $\xi = 6\ell (K\ell)^2/[1 - (K\ell)^4]$.

\begin{figure*}[t]
\includegraphics[width=1\columnwidth]{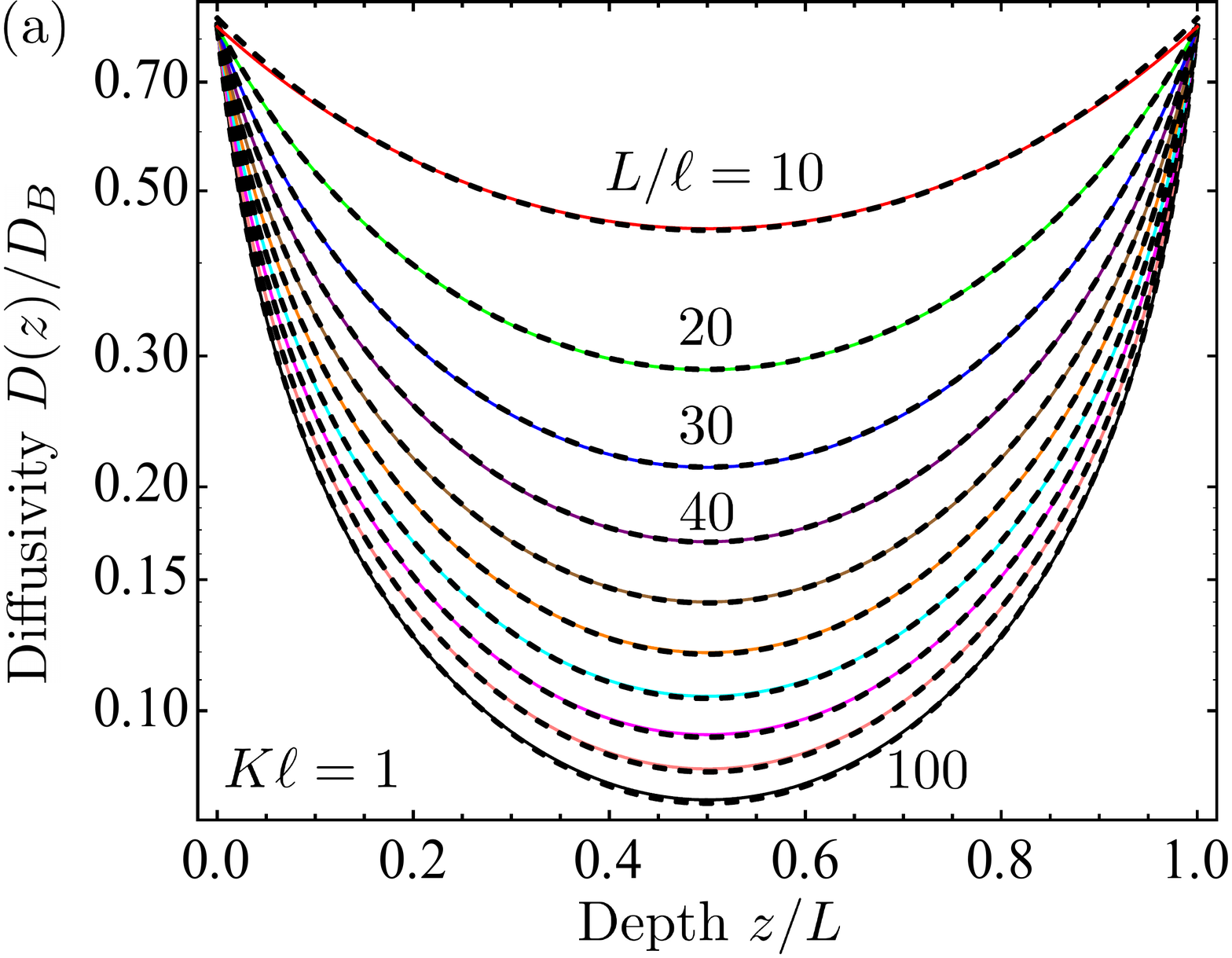}
\includegraphics[width=1\columnwidth]{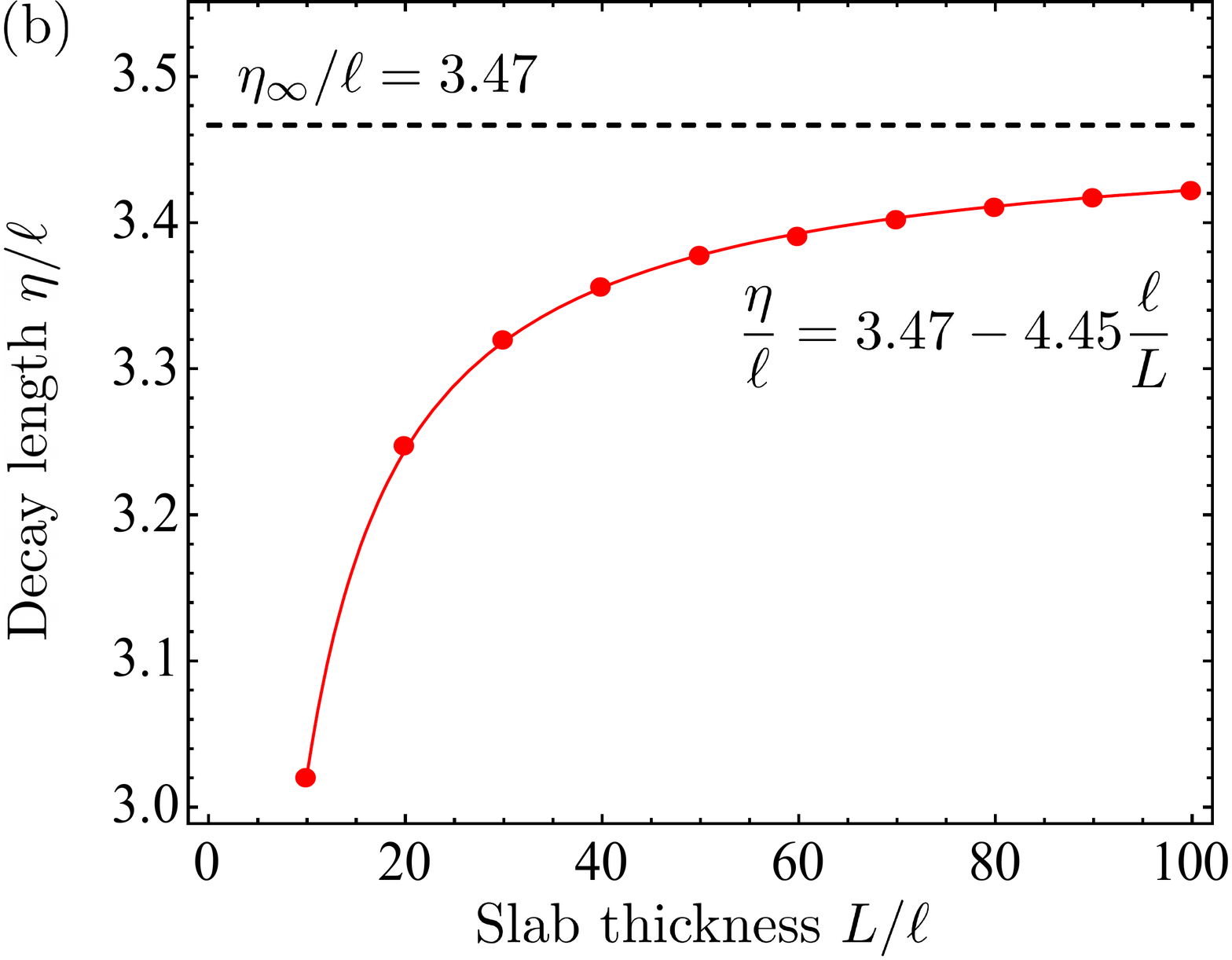}
\caption{\label{fig_dz_me}
Position dependence of the diffusivity $D(z)$ in disordered slabs of different thicknesses $L$ at the mobility edge (a). Solid lines show the numerical solutions of Eqs.\ (\ref{sc1})--(\ref{bc}). Dashed lines show fits of Eq.\ (\ref{dzme}) to numerical results with $\eta$ as a free fit parameter. The best-fit values of $\eta$ are shown in the panel (b) together with a power-law fit that describes the convergence of $\eta$ to a constant value $\eta_{\infty} \simeq 3.47$ with increasing $L$.}
\end{figure*}

For a disordered, infinitely wide slab confined between the planes $z=0$ and $z=L$, $D(\vec{r}) = D(z)$ and Eqs.\ (\ref{sc1}) and (\ref{sc2}) should be supplemented with boundary conditions \cite{tiggelen00sm}
\begin{eqnarray}
&& P(\vec{r},\vec{r}^{\prime}) \mp z_0 \frac{D(z)}{D_B}
\frac{\partial}{\partial z} P(\vec{r},\vec{r}^{\prime})=0
\label{bc}
\end{eqnarray}
at the slab boundaries $z = 0$ and $z = L$, respectively. We use $z_0 = 2\ell/3$ corresponding to the absence of internal reflections \cite{cherroret08sm}.

To solve Eqs.\ (\ref{sc1})--(\ref{bc}) numerically in a slab of disordered medium, we follow an iteration algorithm detailed in Ref.\ \cite{cobus18sm}. In short, we start from $D(z) = D_B$, solve the Fourier transform of Eq.\ (\ref{sc1}) for ${\hat P}(q_ {\perp}, z)$, compute $P(\vec{r},\vec{r})$ using Eq.\ (\ref{ft2}) and then $D(z)$ using Eq.\ (\ref{sc2}). The whole procedure is then repeated many times until convergence of $D(z)$ to a stable profile that does not change from iteration to iteration anymore.

Figure \ref{fig_dz_loc} shows $D(z)$ obtained by solving Eqs.\ (\ref{sc1})--(\ref{bc}) numerically for slabs of different thicknesses $L$ in the localized regime. We have set $K\ell = 0.55$ resulting in the localization length $\xi = 2 \ell$ but similar results are obtained for all $K\ell < 1$ as far as $L$ is much larger than $\xi$. We compare our numerical solution to a simple analytic interpolation formula
\begin{eqnarray}
D(z) =
\left[ D_{\infty}(z)^{1/2} + D_{\infty}(L-z)^{1/2} \right]^2,
\label{dzloc}
\end{eqnarray}
where
\begin{eqnarray}
D_{\infty}(z) = D_{\infty}(0) \exp(-2 z/\xi)
\label{dinfloc}
\end{eqnarray}
is the solution of Eqs.\ (\ref{sc1})--(\ref{bc}) with the same $K\ell < 1$ but in the semi-infinite medium ($L \to \infty$). We find that the agreement of Eq.\ (\ref{dzloc}) with numerical results is remarkable not only for the parameters used in Fig.\ \ref{fig_dz_loc} but also for all $L \gg \xi > \ell$. This suggests that this equation can be used as a reliable analytic approximation of numerical results provided that the condition $L \gg \xi > \ell$ is obeyed.

The behavior of $D(z)$ at the mobility edge $K\ell = 1$ can also be described by a simple interpolation formula similar (but not identical) to Eq.\ (\ref{dzloc}). We find that
\begin{eqnarray}
D(z) = \left[ D_{\infty}(z)^2 + D_{\infty}(L-z)^2 \right]^{1/2}
\label{dzme}
\end{eqnarray}
with
\begin{eqnarray}
D_{\infty}(z) = D_{\infty}(0)
\frac{1}{1 + z/\eta}
\label{dinfme}
\end{eqnarray}
as established previously \cite{tiggelen00sm}, provides very good fits to the numerical solutions of Eqs.\ (\ref{sc1})--(\ref{bc}), see Fig.\ \ref{fig_dz_me}. The fit parameter $\eta$ is a length scale of the order of the mean free path. It slowly converges to roughly $3.5 \ell$ when the thickness of the slab $L$ increases, as we illustrate in Fig.\ \ref{fig_dz_me}(b). This justifies our use of Eq.\ (\ref{dzme}) as an analytic approximation for $D(z)$ at the mobility edge, in the main text.

\begin{figure*}[t]
\includegraphics[width=1\columnwidth]{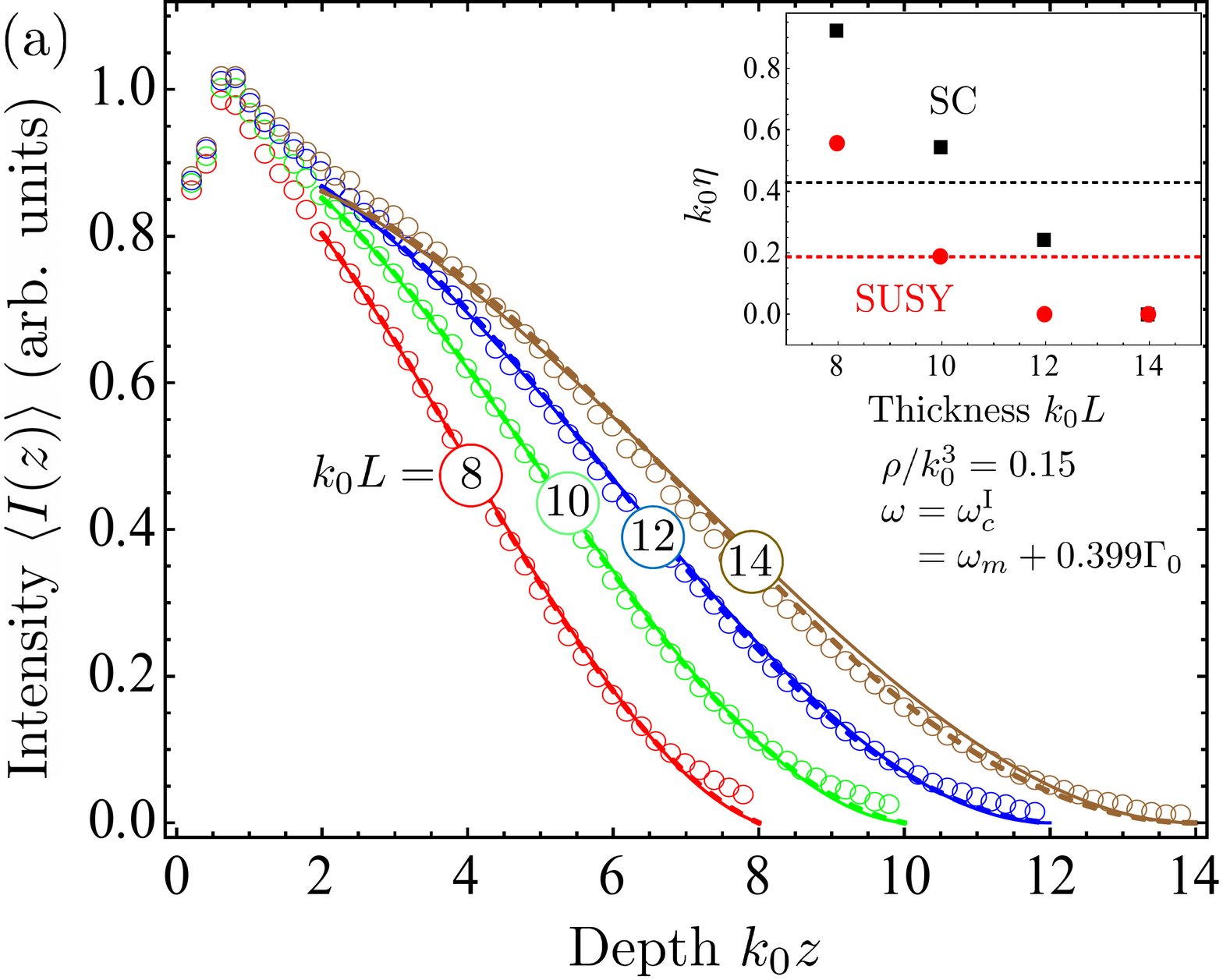}
\includegraphics[width=1\columnwidth]{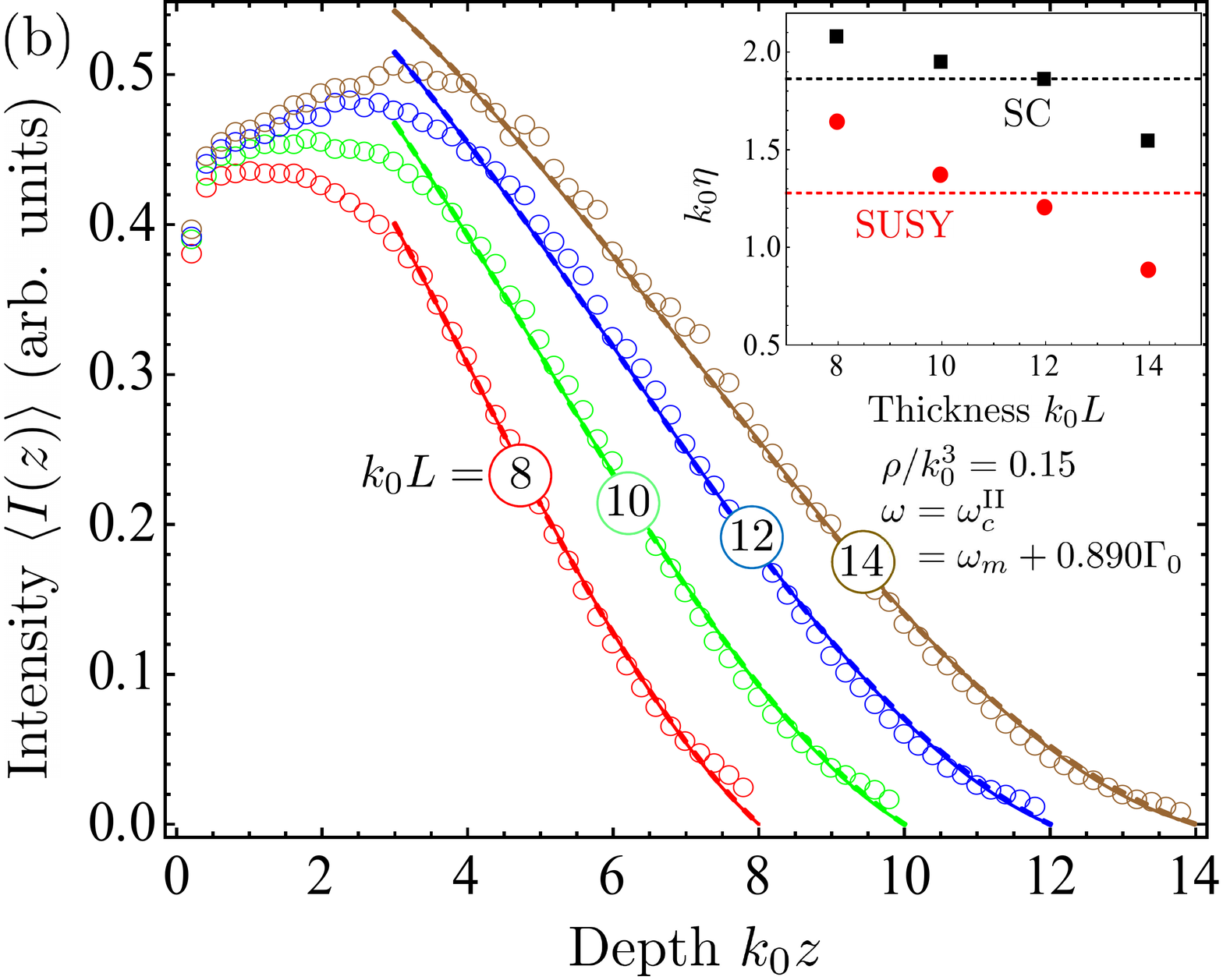}
\caption{\label{fig_me_mf}
Same as Fig.\ 2 of the main text but for scattering of circularly polarized light by two-level atoms in a strong magnetic field $\Delta = 10^3$.}
\end{figure*}

Substituting Eq.\ (\ref{dzme}) into Eq.\ (6) of the main text yields
\begin{eqnarray}
\langle I(z) \rangle_{\mathrm{SC}} &=& I_0 \left( 1 +
\left\{ \vphantom{ \ln \left(\frac{L-\sqrt{2} \eta  u}{L-\sqrt{2} \eta  w-2 z}\right)}
\left[L^3 (u-w)
\right. \right. \right.
\nonumber \\
&+& \left. \left. \left. 2 L^2 (\eta  u-2 u z-\eta  w)
\right. \right. \right.
\nonumber \\
&+& \left. \left. \left. 2 L \left(\eta ^2 (u-w)+3 u z^2-2 \eta  u z\right)
\right. \right. \right.
\nonumber \\
&-& \left. \left. \left. 4 u z \left(\eta
   ^2+z^2\right)\right]-(3/\sqrt{2}) \eta  u w (2 \eta +L)^2
\right. \right.
\nonumber \\
&\times& \left. \left. \ln \left(\frac{L-\sqrt{2} \eta  u}{L-\sqrt{2} \eta  w-2
   z}\right)
\right\} \right.
\nonumber \\
&\times&
\left. \left\{ \vphantom{\ln \left(\frac{\sqrt{2} \eta  u-L}{2 \eta+L}\right)}
\eta  u w \left[2 \eta  L u+3 \sqrt{2} (2 \eta +L)^2
\right. \right. \right.
\nonumber \\
&\times& \left. \left. \left. \ln \left(\frac{\sqrt{2} \eta  u-L}{2 \eta
   +L}\right)\right]
\right\}^{-1}  \right),
   \label{imesc}
\end{eqnarray}
where $u = [2+L/\eta (2+L/\eta)]^{1/2}$ and $w = \{ 2+L/\eta [2+L/\eta (1+2 (L/\eta-1) L/\eta) ] \}^{1/2}$.
We use this expression to fit the results of numerical simulations in Figs.\ 2 and \ref{fig_me_mf}.

\vspace{5mm}
\noindent
\textbf{Light scattering by atoms in magnetic field.}
To compute the spatial distribution of light intensity in a slab filled with two-level atoms subjected to an external magnetic field $\vec{B}$ parallel to the $z$ axis of the coordinate system, we apply a combination of approaches developed previously in Refs.\ \cite{fofanov13sm}, \cite{skip16prasm} and \cite{skip19sm}. The interaction of light with $N$ identical two-level atoms (resonance frequency $\omega_0$, natural decay rate of the excited state $\Gamma_0$, total angular momenta of the ground and excited states $J_g = 0$ and $J_e = 1$, respectively) located at random positions $\{ \vec{r}_j\}$, $j = 1, \ldots, N$, is described by the following Hamiltonian:
\begin{eqnarray}
{\hat H} &=& \sum\limits_{j=1}^{N} \sum\limits_{m=-1}^{1} \left(
\hbar \omega_0 + g_{e} \mu_{\mathrm{B}} B m \right) | e_{jm} \rangle
\langle e_{jm}|
\nonumber \\
&+&
\sum\limits_{\mathbf{s} \perp \mathbf{k}} \hbar ck
\left( {\hat a}_{\mathbf{k} \mathbf{s}}^{\dagger} {\hat a}_{\mathbf{k}\mathbf{s}} + \frac12 \right)
- \sum\limits_{j=1}^{N} {\hat{\vec{D}}}_j
\cdot {\hat{\vec{E}}}(\mathbf{r}_j)
\nonumber \\
&+& \frac{1}{2 \epsilon_0}
\sum\limits_{n \ne j}^{N} {\hat{\vec{D}}}_j \cdot {\hat{\vec{D}}
}_n \delta(\mathbf{r}_j - \mathbf{r}_n),
\label{ham}
\end{eqnarray}
where ${\hat{\vec{D}}}_j$ are the atomic dipole operators, ${\hat{\vec{E}}}(\mathbf{r}_j)$ is the electric displacement vector divided by the vacuum permittivity $\epsilon_0$, ${\hat a}_{\mathbf{k} \mathbf{s}}^{\dagger}$ and ${\hat a}_{\mathbf{k}\mathbf{s}}$ are the photon creation and annihilation operators corresponding to a mode of the free electromagnetic field having a wave vector  $\vec{k}$ and a polarization $\vec{s}$, $2\pi\hbar$ is the Planck's constant, $\mu_{\mathrm{B}}$ is the Bohr magneton, and $g_{e}$ is the Land\'{e} factor of the excited state.
$| e_{jm} \rangle$ denotes the excited state of the atom $j$ in which the eigenvalue of the projection of the total angular momentum operator $\hat{\vec{J}}_e$ on the quantization axis $z$ is equal to $m$.

Following our previous work \cite{skip16prasm}, we introduce a $3N \times 3N$ Green's matrix $G$ of the considered spatial configuration $\{ \vec{r}_j \}$ of atoms:
\begin{eqnarray}
G_{e_{j m} e_{n m'}} &=& \left(i - 2 m \Delta \right) \delta_{e_{j m} e_{n m'}} +
\frac{2 k_0^3}{\hbar \Gamma_0} (1 - \delta_{e_{j m} e_{n m'}})
\nonumber \\
&\times&
\sum\limits_{\mu, \nu}
{d}_{e_{j m} g_j}^{\mu} {d}_{g_n e_{n m'}}^{\nu}
\frac{e^{i k_0 r_{jn}}}{k_0 r_{jn}}
\nonumber
\\
&\times& \left[
\vphantom{\frac{r_{jn}^{\mu} r_{jn}^{\nu}}{r_{jn}^2}}
 \delta_{\mu \nu}
P(i k_0 r_{jn})
+ \frac{r_{jn}^{\mu} r_{jn}^{\nu}}{r_{jn}^2}
Q(i k_0 r_{jn})
\right].
\label{green}
\end{eqnarray}
Here $P(x) = 1 - 1/x + 1/x^2$, $Q(x) = -1 + 3/x - 3/x^2$, $k_0 = \omega_0/c$, $\Delta = g_{e} \mu_{\mathrm{B}} B/\hbar\Gamma_0$ is the Zeeman shift in units of $\Gamma_0$, $\vec{d}_{e_{i m} g_i} = \langle J_{e} m|{\hat{\vec{D}}}_i | J_{g} 0 \rangle$, and $\vec{r}_{jn} = \vec{r}_j - \vec{r}_n$.
The resolvent
\begin{eqnarray}
{\cal R}(\omega) = \left[(\omega-\omega_0) \mathbb{1} + (\Gamma_0/2) G \right]^{-1},
\label{resolvent}
\end{eqnarray}
allows us to compute the position-dependent population $P(\vec{r})$ of excited states. We assume that the atomic sample has a shape of cylinder of radius $R$ and length (thickness) $L \ll R$ parallel to the $z$ axis, and that it is illuminated by a circularly polarized monochromatic plane wave with a frequency $\omega$, a wave vector $\vec{k_{\mathrm{in}}} = (\omega/c) \vec{e}_z$: $\mathbf{E}_{\mathrm{in}}(\mathbf{r}) = \mathbf{u}_{\mathrm{in}} E_0 \exp(i \mathbf{k}_{\mathrm{in}} \mathbf{r})$, where the unit vector $\mathbf{u}_{\mathrm{in}}$ ($|\mathbf{u}_{\mathrm{in}}| = 1$) determines the polarization of the incident light. Then \cite{fofanov13sm}
\begin{eqnarray}
P(\vec{r})  &=& \lim\limits_{\Delta V \to 0} \frac{1}{\hbar^2 \Delta V}
\sum\limits_{m=-1}^1
\left| \sum\limits_{\vec{r}_j \in \Delta V}
\sum\limits_{n, m'} {\cal R}_{e_{jm} e_{nm'}} \right.
\nonumber \\
&\times& \left. \vec{d}_{e_{n m'} g_n} \cdot \vec{E}_{\mathrm{in}}(\vec{r}_n)
\vphantom{\sum\limits_{\vec{r}_j \in \Delta V}}
\right|^2,
\label{popm}
\end{eqnarray}
where $\Delta V$ is an infinitesimal volume element centered at $\vec{r}$.
We now average $P(\vec{r})$ over a wide circular area of radius $R_1 < R$ around the cylinder axis and over a large number (up to $9 \times 10^5$) statistically independent atomic configurations $\{ \vec{r}_j \}$. This yields a quantity that depends only on $z$ and not on $\vec{r}_{\perp} = \{x, y \}$ and approximates the population of excited states in an infinitely wide ($R \to \infty$) slab of atoms:
\begin{eqnarray}
&&\langle P(z) \rangle = \frac{1}{\pi R_1^2}
 \int\limits_{r_{\perp} < R_1 < R} \langle P(\vec{r} = \left\{ \vec{r}_{\perp}, z \right\}) \rangle d^2 \vec{r}_{\perp}.\;\;\;\;\;
\label{popm1}
\end{eqnarray}
Variance of $P(\vec{r})$ can be calculated in the same way, by averaging $[P(\vec{r}) -  \langle P(z) \rangle]^2$ instead of $P(\vec{r})$.

Finally, we assume that the population of excited states $P(\vec{r})$ and the intensity of light $I(\vec{r})$ are proportional to each other with a proportionality coefficient that is independent of position $\vec{r}$. This assumption is justified by the linearity of the considered physical system and allows us to analyze $P(\vec{r})$ instead of $I(\vec{r})$ and vice versa, as far as the we are not interested in the absolute magnitude of $P$ or $I$ but only in their variations in space.

\begin{figure}[t]
\includegraphics[width=1\columnwidth]{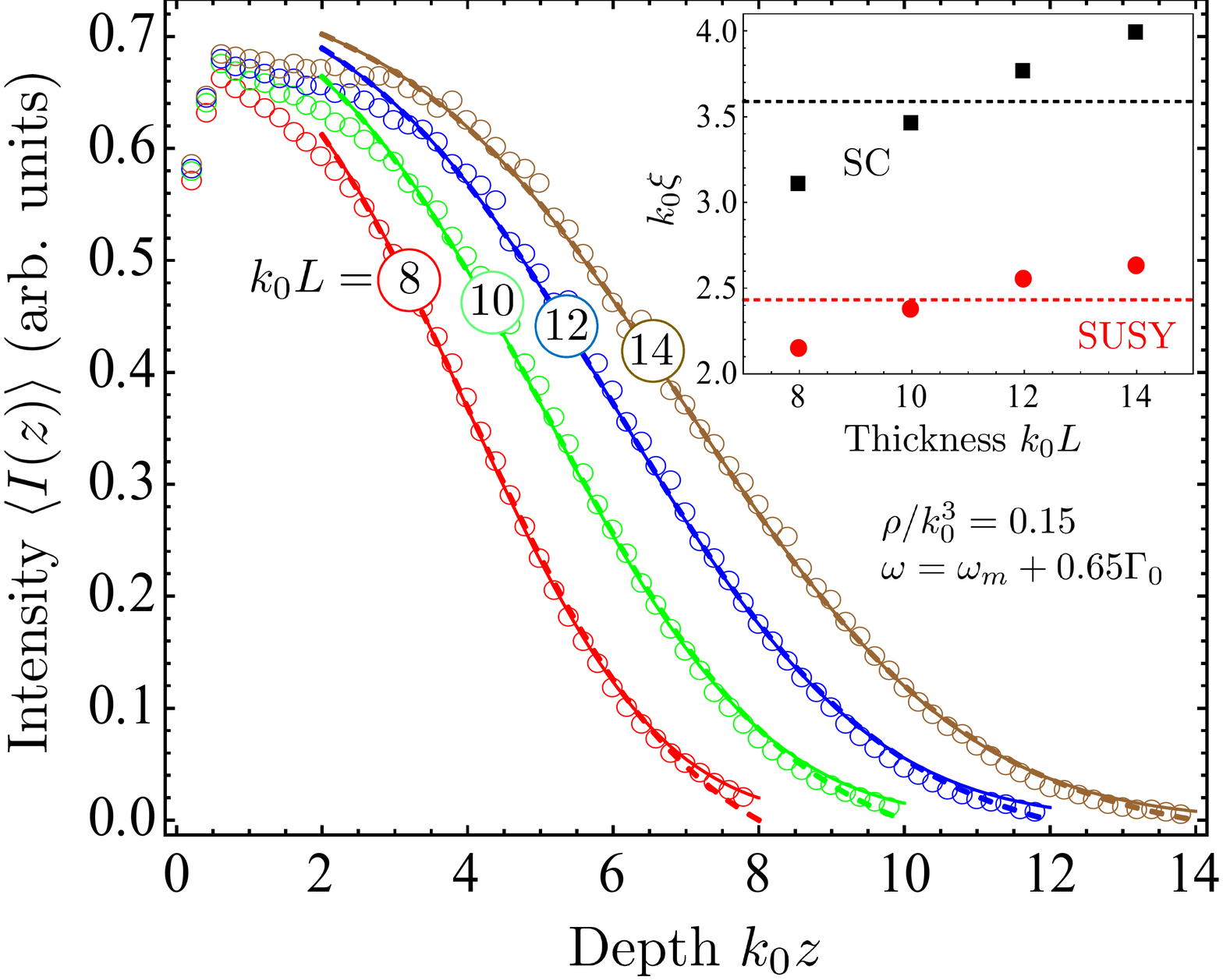}
\caption{\label{fig_loc_mf}
Same as Fig.\ 1 of the main text but for scattering of circularly polarized light by two-level atoms in a strong magnetic field $\Delta = 10^3$.}
\end{figure}

\begin{figure}
\hspace*{-6mm}
\includegraphics[width=1.1\columnwidth]{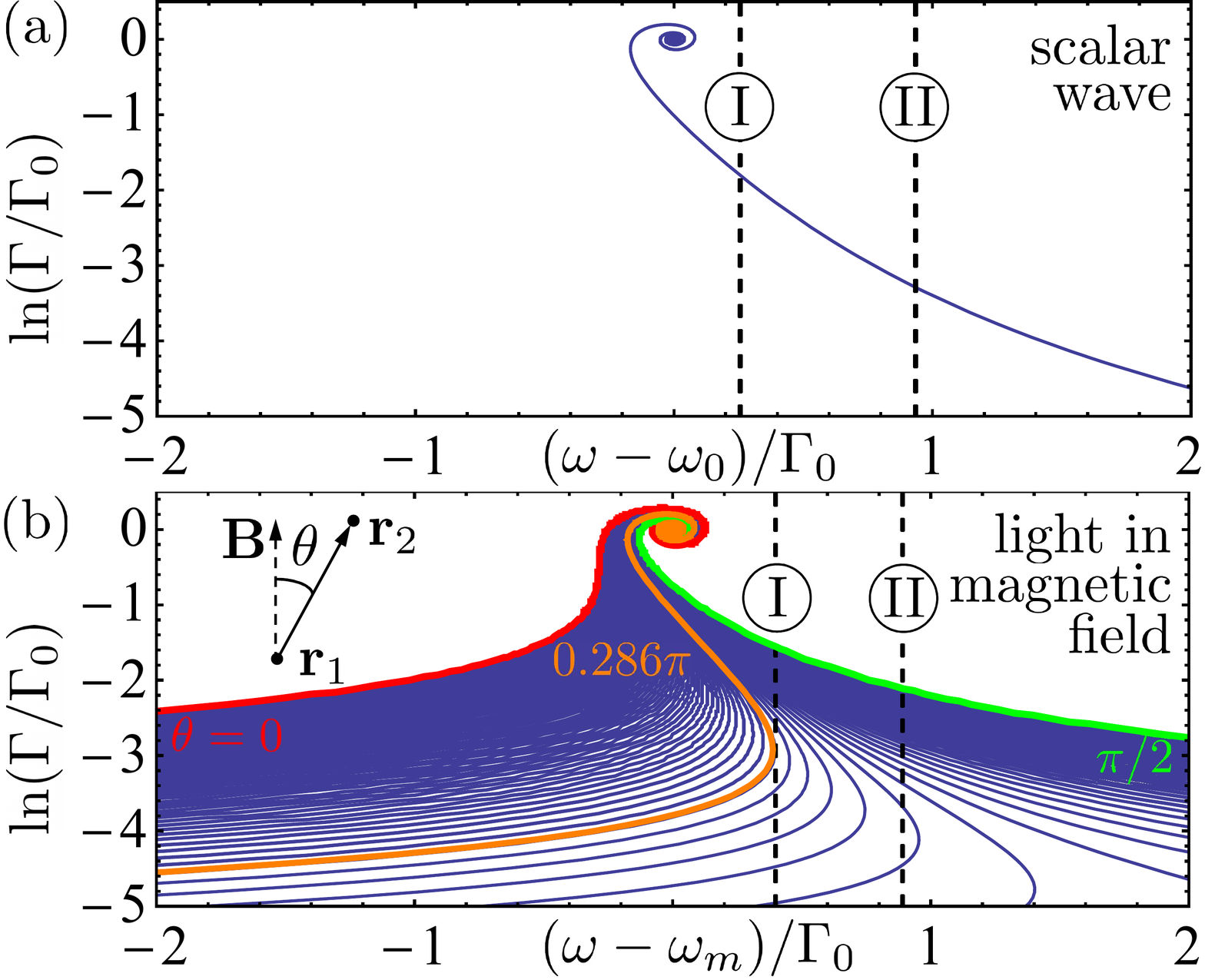}
\caption{\label{fig_2atoms}
Comparison of trajectories followed by eigenvalues $\Lambda$ of the Green's matrix ${\hat G}$ on the complex plane  $(\omega -\omega_m)/\Gamma_0 = - \mathrm{Re}\Lambda/2$, $\Gamma/\Gamma_0 = \mathrm{Im}\Lambda$ for $N = 2$ scatterers (atoms) in the scalar case [(a), $m = 0$] and for light in magnetic field [(b), $m = \pm 1$]. Eigenvalues move away from the point ($\omega = \omega_m$, $\Gamma = \Gamma_0$) along the lines shown in the figure as the distance between the scatterers $\Delta r$ decreases. In (b), the trajectory followed by the eigenvalue depends on the angle $\theta$ between $\Delta \vec{r} = \vec{r_2}-\vec{r}_1$ and the magnetic field $\vec{B}$; all possible trajectories are comprised between two limiting curves corresponding to $\theta = 0$ (shown in red) and $\theta = \pi/2$ (shown in green). Trajectories reach the frequency band of localized states only for $\theta \gtrsim 0.286 \pi$ (shown in orange). Only eigenvalues corresponding to subradiant states with $\Gamma < \Gamma_0$ for $\Delta r \to 0$ are shown. Vertical dashed lines show locations of mobility edges [same as in Fig.\ 3(b) for (a) and Fig.\ \ref{fig_fluct_mf} for (b)].}
\end{figure}

In order to determine the positions of mobility edges in the considered scattering system, we rely on the previous work \cite{skip18prlsm} where a localization phase diagram has been established for it in the limit of $B \to \infty$. In particular, we have found that four mobility edges exist when the atomic number density $\rho$ is high enough. They form closely located pairs near frequencies $\omega_m = \omega_0 + m \Gamma_0 \Delta$, $m = \pm 1$. We choose one such a pair $\omega_c^{\mathrm{I}} = \omega_m + 0.399 \Gamma_0$ and $\omega_c^{\mathrm{II}} = \omega_m + 0.890 \Gamma_0$ for  $\rho/k_0^3 = 0.15$. A frequency $\omega$ in between $\omega_c^{\mathrm{I}}$ and $\omega_c^{\mathrm{II}}$ corresponds to Anderson localized states. Average intensity profiles obtained for a circularly polarized incident plane wave [$\vec{u}_{\mathrm{in}} = (\vec{e}_x \pm i\vec{e}_y)/\sqrt{2}$ for $m = \mp 1$] at such a frequency are shown in Fig.\ \ref{fig_loc_mf} (symbols) together with SC and SUSY theory fits. We see that the agreement between numerical simulations and theory is as good as in the scalar case (cf. Fig.\ 1 of the main text) and that the best-fit $\xi$ of SUSY theory vary less with $k_0 L$ than $\xi$ of SC theory, showing a tendency to saturation at large $k_0 L$. The agreement between numerical simulations and the local diffusion theories remains good at the two mobility edges, see Fig.\ \ref{fig_me_mf}. This allows us to conclude that the analytical expressions for $\langle I(z) \rangle$ derived in this work are valid beyond the scalar wave model and are likely to be universal for Anderson localization of any waves.

\vspace{5mm}
\noindent
\textbf{Role of two-atom subradiant states.}
A difference between the two mobility edges of the scalar model discussed in the main text is attributed to subradiant states localized on pairs of closely located atoms. Here we discuss this issue in more details.

For a scattering system composed of $N = 2$ atoms separated by a distance $\Delta r$, the two eigenvalues of the $2 \times 2$ matrix ${\hat G}(\omega_0)$ are
\begin{eqnarray}
\label{ev2}
\Lambda_{\pm} = i \pm \frac{\exp(i k_0 \Delta r)}{k_0 \Delta r},
\end{eqnarray}
whereas the eigenvectors are
\begin{eqnarray}
\label{evect2}
\bm{\psi}_{\pm} = \frac{1}{\sqrt{2}} \left\{ 1, \pm 1 \right\}^{T}.
\end{eqnarray}

The eigenvector $\bm{\psi}_{+}$ describes a \textit{superradiant} quasimode with a decay rate $\Gamma_+ = \Gamma_0 \mathrm{Im} \Lambda_+ > \Gamma_0$. It is strongly sensitive to the surroundings of the two-atom system and will be strongly modified when more atoms are added, in particular when $k_0 \Delta r \gtrsim 1$. For $k_0 \Delta r < 1$, its large negative frequency detuning $(\omega-\omega_0)/\Gamma_0 = -\frac12 \mathrm{Re} \Lambda_+ < 0$ makes its traces surviving in an ensemble of many atoms irrelevant for the analysis of Anderson localization that we find to occur for $(\omega-\omega_0)/\Gamma_0 > 0$. In contrast, the \textit{subradiant} quasimode $\bm{\psi}_{-}$ has a positive frequency detuning $(\omega-\omega_0)/\Gamma_0 = -\frac12 \mathrm{Re} \Lambda_- > 0$ for $k_0 \Delta r < 1$ and a decay rate $\Gamma_- = \Gamma_0 \mathrm{Im} \Lambda_- < \Gamma_0$. It is weakly sensitive to other atoms put near the two-atom system. Figure\ \ref{fig_2atoms}(a) shows the trajectory followed by the subradiant eigenvalue on the complex plane $(\omega -\omega_0)/\Gamma_0 = - \mathrm{Re}\Lambda/2$, $\Gamma/\Gamma_0 = \mathrm{Im}\Lambda$ as $\Delta r$ decreases. The trajectory crosses the first mobility edge for $k_0 \Delta r \simeq 1$, which corresponds to weakly bound pairs that do not survive in the ensemble of many atoms as follows from Fig.\ 3(b) where no strongly localized states (i.e., black points) exist on the left from the mobility edge I. Therefore, the properties of the mobility edge I are not significantly affected by two-atom states. In contrast, the second mobility edge corresponds to  $k_0 \Delta r \simeq 0.5$ which turns out to be small enough to resist the influence of other atoms. As a result we see states localized on pairs of atoms and having $\mathrm{IPR} \simeq 0.5$ in the immediate vicinity of the mobility edge II. These states coexist with extended states [light grey points in Fig.\ 3(a)] and lead to `giant' fluctuations of intensity at the second mobility edge.

\begin{figure}[t!]
\hspace*{-6mm}
\includegraphics[width=1.1\columnwidth]{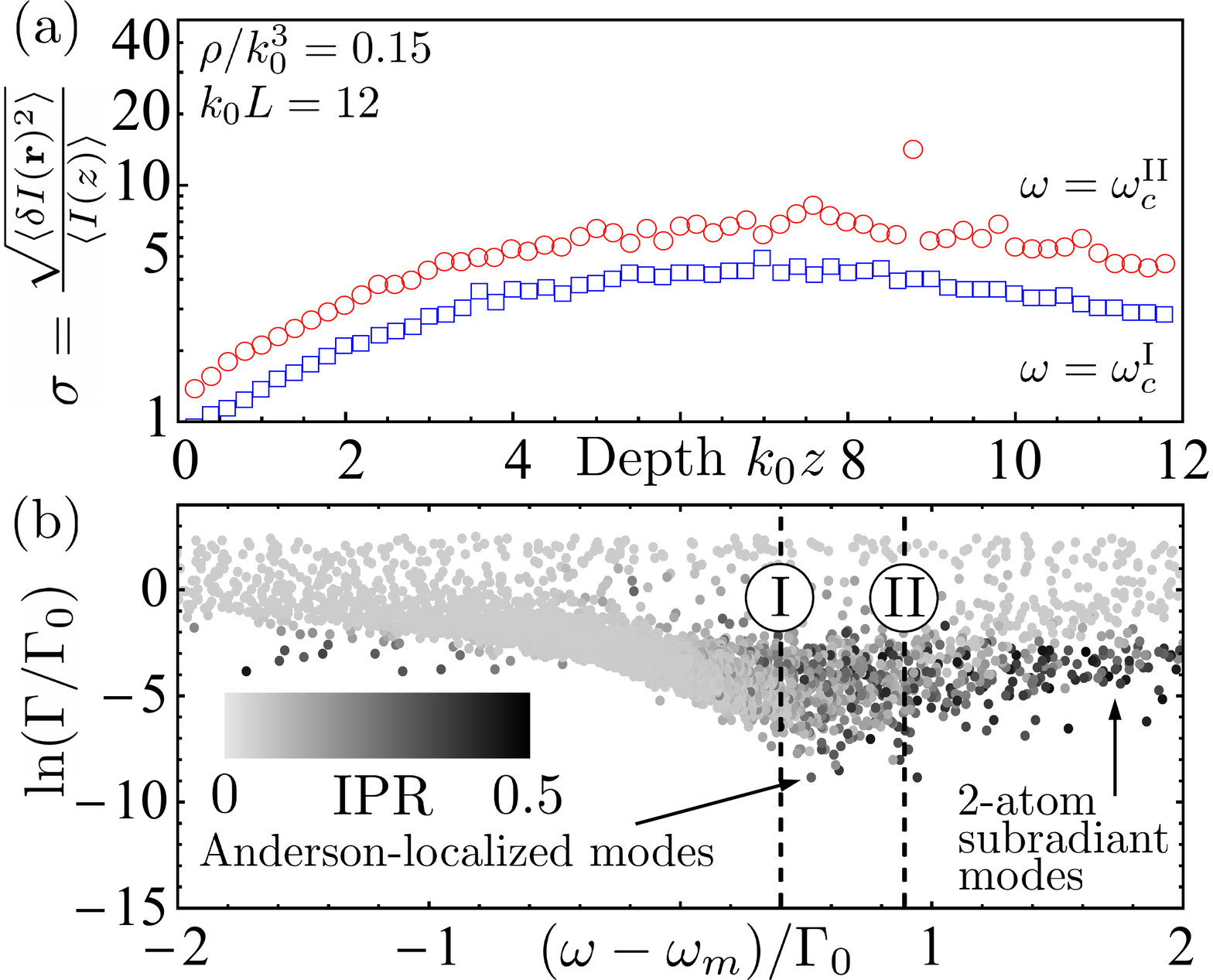}
\caption{\label{fig_fluct_mf}
Same as Fig.\ 3 of the main text but for scattering of circularly polarized light by two-level atoms in a strong magnetic field $\Delta = 10^3$. Averaging is performed over $7 \times 10^5$ independent random atomic configurations.}
\end{figure}

An indirect confirmation of the impact of two-atom subradiant states on the magnitude of intensity fluctuations at the high-frequency mobility edge can be obtained by analyzing the case of light scattering by atoms subjected to a strong external magnetic filed (see the previous section of this Supplemental Material). An explicit expression for the eigenvalues and eigenvectors of the matrix ${\hat G}(\omega_0)$ can be readily obtained by diagonalizing the matrix (\ref{green}) for $N = 2$. Now the trajectory followed by the subradiant eigenvalue upon decreasing the distance $\Delta r$ between the two atoms depends on the angle $\theta$ between the vector $\Delta \vec{r} = \vec{r}_2 - \vec{r}_1$ and the external magnetic field $\vec{B}$, see Fig.\ \ref{fig_2atoms}(b). Possible trajectories fill a large part of the complex plane. For $\theta \lesssim 0.286 \pi$, they do not reach the mobility edges, remaining on the left from the mobility edge I for any $\Delta r$. In addition, the distances $\Delta r$ corresponding to crossings between subradiant eigenvalues and mobility edges are larger than in the scalar case for almost all other values of $\theta$. Therefore, the impact of two-atom states on the quantities calculated near mobility edges and in the Anderson localized regime is less important than in the scalar case, and the system is expected to exhibit similar behaviors at both mobility edges. In particular, the variance of intensity fluctuations at the second mobility edge exceeds its values at the first one only slightly, see Fig.\ \ref{fig_fluct_mf}(a). This is in contrast with the result obtained for scalar waves [see Fig.\ 3(a)] but correlates with less pronounced two-atoms states in Fig.\ \ref{fig_fluct_mf}(b) as compared to Fig.\ 3(b). Indeed, IPR of eigenvalues appearing on the right from the mobility edge II in Fig.\ \ref{fig_fluct_mf}(b) is smaller [the points are lighter than in Fig.\ 3(b)] and `dark' eigenvalues corresponding to localized states are less separated from `light' ones (extended states) in the vertical direction, having similar lifetimes. This confirms the correlation between stronger intensity fluctuations at the second mobility edge and subradiant states localized on pairs of closely located atoms.

Finally, we note that fluctuations of intensity of scattered light near mobility edges and in the localized regime have been recently studied by Cottier \textit{et al.} \cite{cottier19sm}. These authors calculated intensity variance for light outside a disordered medium but some of their observations are similar to ours: stronger intensity fluctuations at the high-frequency mobility edge in the scalar model (see their Fig.\ 3) and apparent differences between the scalar case and the case of light scattering by atoms in a strong magnetic filed (compare their Figs.\ 3 and 4). Our analysis presented above suggests that those observations can also be explained by two-atom subradiant states that survive multiple scattering and have different properties in different models.

\end{document}